\documentclass[aps,twocolumn,superscriptaddress,longbibliography]{revtex4-1}

\usepackage{graphicx}% Include figure files
\usepackage{dcolumn}% Align table columns on decimal point
\usepackage{bm}% bold math
\usepackage{amsfonts}
\usepackage{amsmath}
\usepackage{amssymb}
\usepackage{xcolor}
\setcitestyle{super}
\usepackage{enumerate}
\usepackage{soul}

\usepackage{hyperref}% add hypertext capabilities
\hypersetup{
    colorlinks,%
    citecolor=blue,%
    linkcolor=blue,%
    urlcolor=blue
}

\newcommand{\bl}{BBL($\alpha^\prime$)$_\text{1/9}$}

\newcommand{\bla}{BBL$\text{(a)}_{1/4}$}
\newcommand{\blb}{BBL$\text{(b)}_{1/4}$}

\newcommand{\orcid}[1]{\href{https://orcid.org/#1}{\includegraphics[width=8pt]{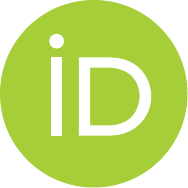}}}

\begin{document}

\title{Bridging Borophene and Metal Surfaces: \\ Structural, Electronic, and Electron Transport Properties}

\author{Wanderl\~a L. Scopel\orcid{0000-0002-2091-8121}} 
\email{wanderla.scopel@ufes.br}
\affiliation{Physics Department, Federal University of Espírito Santo, 29075-910, Vitória, ES, Brazil}

\author{F. Crasto de Lima\orcid{0000-0002-2937-2620}} 
\email{felipe.lima@ilum.cnpem.br}
\affiliation{Ilum School of Science, CNPEM, 13083-970, Campinas, SP, Brazil}

\author{Pedro H. Souza\orcid{0000-0001-7009-7379}} 
\email{psouza8628@gmail.com}
\affiliation{Instituto de F\'isica, Universidade Federal de Uberl\^andia, 38400-902, Uberl\^andia, MG, Brazil}

\author{Jos\'e E. Padilha\orcid{0000-0002-4299-2808}} 
\email{jose.padilha@ufpr.br}
\affiliation{Campus Avan\c{c}ado Jandaia do Sul, Universidade Federal do Paran\'a, 86900-000, Jandaia do Sul, PR, Brazil}

\author{Roberto H. Miwa\orcid{0000-0002-1237-1525}} 
\email{hiroki@ufu.br}
\affiliation{Instituto de F\'isica, Universidade Federal de Uberl\^andia, 38400-902, Uberl\^andia, MG, Brazil}

\date{\today}

\begin{abstract}

Currently, solid interfaces composed of two-dimensional materials (2D) in contact with metal surfaces (m-surf)  have been the subject of intense research, where the borophene bilayer (BBL) has been considered a prominent material for the development of electronic devices based on 2D platforms.  In this work, we present a theoretical study of the energetic, structural, and electronic properties of the BBL/m-surf interface, with m-surf = Ag, Au, and Al (111) surfaces, and the electronic transport properties of BBL channels connected to the BBL/m-surf top contacts.  We find that the bottom-most BBL layer becomes metalized, due to the orbital hybridization with the metal surface states, resulting in BBL/m-surf ohmic contacts, meanwhile, the inner and top-most boron layers kept their semiconducting character. The net charge transfers reveal that BBL has become $n$-type ($p$-type) doped for m-surf = Ag, and Al (= Au). A thorough structural characterization of the BBL/m-surf interface, using a series of simulations of the X-ray photoelectron spectra, shows that  the formation of BBL/m-surf interface is characterized by a redshift of the B-$1s$ spectra. Further electronic transport results revealed the emergence of a Schottky barrier between 0.1 and 0.2\,eV between the BBL/m-surf contact and the BBL channels. We believe that our findings are timely, bringing important contributions to the applicability of borophene bilayers for developing 2D electronic devices.

\end{abstract}

\maketitle

\section{Introduction}

Two-dimensional (2D) materials have been intensively studied with a focus on both fundamental issues and practical applications addressing the development of novel electronic devices. Since the successful synthesis of graphene,\cite{novoselovScience2004-2} other 2D systems have been found/proposed, such as single-layer (SL) transition metal dichalcogenides (TMDs), like MoS$_2$ and WSe$_2$ SLs,\cite{radisavljevicNatNanotech2011,allain2015electrical} and graphene counterparts like silicene,\cite{vogtPRL2012} germanene,\cite{bhimanapatiACSNano2015} phosphorene.\cite{liu2014phosphorene} 

Because of its peculiar  electronic structure, boron atoms can exist in a wide range of structural phases, including 2D sheets referred to as borophene.\cite{kunstmannPRB2006,tangPRL2007,penevNanoLett2012,wuACSNano2012,mannixScience2015,fengNatChem2016} In a seminal work, Mannix {\it et al.}\cite{mannixScience2015}  synthesized single-layer borophene on the Ag(111) surface, characterized by a buckled structure of boron atoms forming a triangular lattice. In the sequence, Feng {\it et al.}\cite{fengNatChem2016} synthesized a planar and more stable SL borophene structure distinguished by the formation of periodic holes embedded in a triangular lattice. Other metal surfaces, besides Ag(111),  have also been utilized as a substrate for the  borophene's synthesis, for instance,  Al(111), Au(111), Cu(111), and Ir(111).\cite{kiraly2019borophene,kaneti2021borophene}

Theoretical research on the electronic characteristics of borophene has been stimulated by the polymorphism of 2D systems based on boron atoms. For instance, (i)  exotic electronic properties, like the formation of Dirac nodal lines\cite{gupta2018dirac} and the formation of chiral states in borophene bilayer systems,\cite{crasto2019orbital} among other electronic features. In parallel, (ii) combined with atomic-scale material engineering, borophene heterostructures have been successfully synthesized. For instance, lateral superlattices are composed of (periodic) rows of borophene single layer (BSL) with different structural phases.\cite{liuNatMat2018} In a recent study, we predicted the emergence of electronic confinement effects and anisotropic electronic transport in those borophene lateral superlattices.\cite{padilhaPCCP2016,silvestreNanoscale2019electronic}

Further theoretical studies have predicted the formation of interlayer chemical bonds between SL borophene sheets, resulting in borophene bilayer (BBL) systems.\cite{gao2018structure} Indeed, semiconductor BBLs were produced on metal surfaces like Ag(111) and Cu(111),\cite{liuNatMat2021,xuNanoLett2022quasi,chenNatChem2021}  and subsequent first-principles computations verified their energetic and structural stability.\cite{yang2022bilayer} 

A solid understanding of the structural and electronic properties of new materials is fundamental for the development of electronic devices. Here, computational  simulations, based on first-principles methods, have been viewed as a promising  solution for a comprehensive atomic scale understanding of the electronic and structural properties of materials in general. For instance, in 2D systems like MoS$_2$/metal,\cite{chen2013tuning, gong2014unusual,kang2014computational,parto2021one}  phosphorene/metal,\cite{xu2015electron,pan2016monolayer} and more recently BBL/metal interfaces.\cite{yang2022bilayer,li2022opposite,mozvashi2022effects}

Currently, the applications of new materials in "two-dimensional electronics" relies on critical issues like the understanding of the electronic and structural properties of  metal/semiconductor interfaces  and the emergence of electronic transport  throughout 2D semiconductor channels. \cite{NATELECmay2022,NATELECdas2021,NATshen2021} In this work, we used first-principles calculations to investigate (i) the energetic, structural, and electronic properties of bilayer borophene (BBL) on different metal surfaces (m-surf), specifically Ag(111), Au(111), and Al(111), denoted as BBL/m-surf, and (ii) the electronic transport properties through BBL channels (ch) under different BBL/m-surf top contacts (tc), which we describe as [BBL/m-surf]$_\text{tc}$\,-\,[BBL]$_\text{ch}$\,-\,[BBL/m-surf]$_\text{tc}$. In (i), we focus on the interfacial hybridizations that govern the formation of BBL/m-surf, resulting in either $n$- or $p$-type doping of the BBL, the formation of an ohmic contact at the semiconductor/metal interface, and the m-surf\,$\rightarrow$\,BBL tunneling barrier. In (ii), we calculate the transmission function along the BBL channels, [BBL]$_\text{ch}$, and evaluate the (horizontal) Schottky barrier between the BBL/m-surf]$_\text{tc}$ top-contact  and [BBL]$_\text{ch}$ channel. Our findings reveal that the recently synthesized BBLs may offer suitable electronic properties for the development of electronic devices based on 2D materials.

\section{Computational details}

\subsection{Energetic, structural and electronic properties}

%%%%%%FIG1
 \begin{figure*}
    \includegraphics[width=17cm]{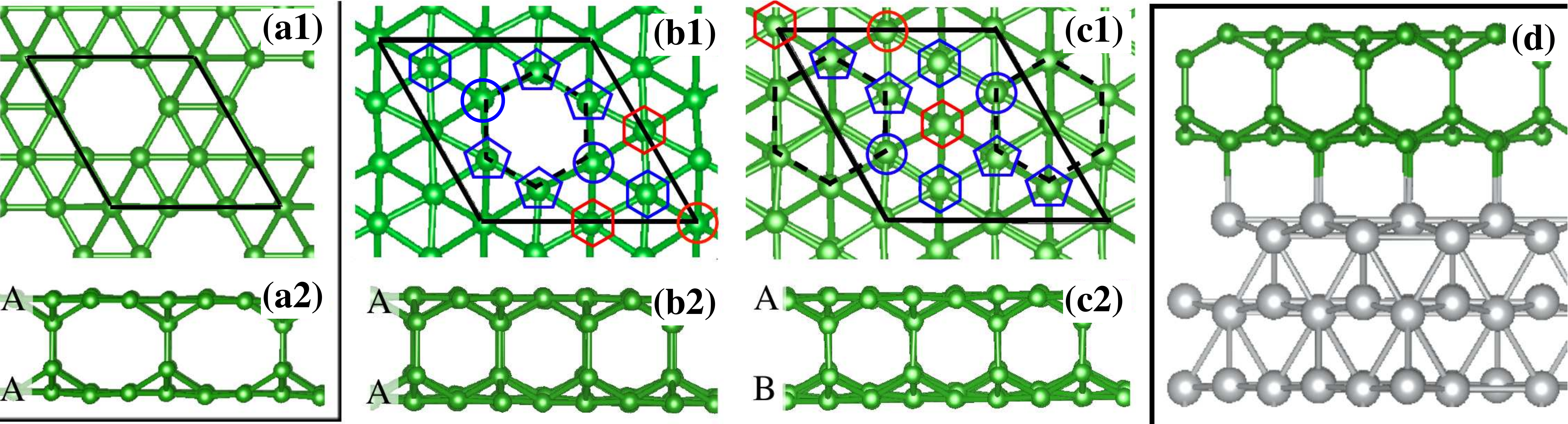}
    \caption{\label{models0} Structural model of borophene bilayer, \bl\, top (a1) and side (a2) views; \bla\, (AA stacking) top (b1) and side (b2) views; \blb\, (AB stacking) top (c1) and side (c2) views. (d) Structural model of BBL adsorbed on the metal (111) surface, BBL/m-surf. Dashed lines in (b1) and (c1) indicate the hollow hexagons in the top layer (A). The non-equivalent six-fold (five-fold) coordinated  atoms are indicated by hexagons (pentagons) and the boron atoms that form interlayer B\,-\,B bonds are indicated by circles, where  different colors indicate different local geometries.}
  \end{figure*}

The calculations of the energetic, structural, and electronic properties were performed by using the density functional theory (DFT) as implemented in the QUANTUM ESPRESSO\cite{espresso} (QE) code. The exchange-correlation energy was obtained within the generalized gradient approximation, as proposed by  Perdew, Burke, and Ernzerhof (GGA-PBE),\cite{PBE} and van der Waals (vdW) interactions were included using the vdW-DF approach.\cite{thonhauserPRB2007, langrethJPhysC2009, thonhauserPRL2015} We have benchmarked our total energy results by using other vdW approaches, vdW-Opt88,\cite{klimesJPhysC2010} -cx,\cite{berland2014exchange,berland2014exchange} and D2.\cite{grimmeJCompChem2006} The borophene monolayer and bilayer systems were described using a supercell with 2$\times$2 surface periodicity, and a vacuum region of 20\,\AA. The metal surfaces were described using a slab with seven metal layers, and  vacuum regions of 35\,\AA\, in order to minimize the long-range interaction between a given surface and its (periodic) image. The Brillouin zone sampling was performed by using a set of 8$\times$8$\times$1 special k-points,\cite{mp} and the Kohn-Sham orbitals were expanded in a plane-wave basis set with an energy cutoff of 60\,Ry, and 480\,Ry for the total charge density. The atomic positions were fully relaxed until the residual forces converged to within 25\,meV/\AA.

\subsection{Electronic transport}

The transport calculation  was carried out with the TranSIESTA code, \cite{transiesta_inicio,transiesta} which combines DFT and NEGF. The core electrons were described by Norm-conserving Troullier-Martins pseudopotentials \cite{ncpseudos}, while the valence electrons were treated with a single-zeta polarized (SZP) basis set of numerical atomic orbitals (NAOs) \cite{numericalatomicorbitals}, with an energy shift of 0.03 Ry.  A real-space mesh cutoff ($300$ Ry) and a k-points mesh  $1\times4\times128$ ($1\times 4\times 1$ )  for the electrodes (scattering) region in reciprocal space based on the Monkhorst-Pack \cite{mp} scheme were implemented. 

For zero-bias calculations, the energy-resolved electronic transmission ( $T(E)$), which represents the probability that an electron from the left electrode reaches the right electrode through the scattering region, is written as:
\begin{equation}
T\left(E \right) = Tr \left[\Gamma_{\mathrm L}\left(E\right) {\cal G}\left(E\right) \Gamma_{\mathrm R}\left(E\right) {\cal G}^{*}\left(E\right) \right]
\label{eq1}
\end{equation}

The coupling matrices are defined as $\Gamma_\alpha = i \left[ \Sigma_\alpha - \Sigma_\alpha^{\dagger} \right]$, with $\alpha\equiv \left\{{\mathrm{L,R}}\right\}$. $\cal G$ ($\cal G ^*$) is the retarded (advanced) Green's function.

The local current of the device is calculated by the Landauer-B\"uttiker formula: \cite{buttiker1984quantum} (regardless electronic spin)
\begin{equation}
  \label{eqI}
  I = \frac{2e}{h}\int_{\mu_R}^{\mu_L}
    T(E)[f(E-\mu_L) - f(E-\mu_R)]dE,
\end{equation}
where $\mu_L$($\mu_R$) represents the electrochemical potential of the left(right) 
electrode, $f(\epsilon)$ is the Fermi-Dirac intercalated distribution for energy 
$\epsilon$, and $T(E)$ is given by Eq.~\ref{eq1}. 

\subsection{Core-level binding energy}

The core-level shift (CLS) energy was calculated within the QE code, with the same special k-points, plane wave basis, and charge density energy cutoff of the energetic, structural, and electronic properties. Here the CLS can be calculated within DFT through the $\Delta SCF$ approach (difference in the self-consistent field), which takes into account the valence electrons relaxation after the core-hole formation, that is, capturing the screening effects \cite{JCPbellafont2015, EJPBgarcia2012}. The binding energy is defined as the difference between the excited state with a core-hole ($E^{(n-1)}$) and the ground state energy ($E^{(n)}$)
    \begin{equation}
        BE = E^{(n-1)} - E^{(n)}.
    \end{equation}
The absolute value of the BE is dependent on the functional choice, however its variation (relative energies) are rather robust \cite{JCPtakahashi2004}, we have taken as reference the experimental BE of h-BN \cite{PCCPrastogi2019}. Our calculations for the BE take the contribution of all B atoms in the unity cell and used a Gaussian width of $0.3$\,eV to construct the spectra.

\section{Results}

We begin by outlining some of the fundamental characteristics of free-standing borophene monolayer (BML) and bilayer (BBL) systems before beginning our study of the BBL on metal surfaces.

\subsection{Free-standing borophene}
  
{\it Monolayer borophene.} The $\alpha$ model for  borophene monolayer, BML($\alpha$), was predicted by Tang {\it et al.},\cite{tangPRL2007,tangPRB2010} as the energetically most stable geometry for single-layer borophene. The BML($\alpha$) phase is characterized by a planar sheet of boron atoms forming a hexagonal pattern of holes with  hole density ($\eta$) of 1/9. Its electronic structure is characterized by fully occupied (unoccupied) in-plane $\sigma$ ($\sigma^\ast$) bonding (anti-bonding) states, combined with partially occupied out-of-plane $\pi$ orbitals. A net charge (re)distribution between the three- and two-center bondings\cite{tangPRL2007} makes the  $\alpha$  phase energetically stable. A few years later, Wu {\it et al.}\cite{wuACSNano2012} verified that such a  planar structure is dynamically unstable; they found imaginary frequencies  in the phonon spectra calculations. However, in the same study, they found that (i) those imaginary frequencies are removed upon  vertical buckling of the boron atoms in an alternated fashion, followed by (ii) a slight increase of the borophene binding energy. Indeed, our results confirm that such a  buckled structure, BML(${\alpha^\prime}$), is energetically more stable than the planar BML($\alpha$) by 5\,meV/B-atom, followed by a nearly negligible reduction  of the  lattice constant (less than 0.12\,\%) when compared with one  of the flat ($\alpha$) phase, namely 5.029\,$\rightarrow$\,5.023\,\AA. In Ref.\,\cite{wuACSNano2012}, based on DFT-PBE calculations, the authors obtained equilibrium bond lengths of 5.050 ($\alpha$) and 5.046\,\AA\,($\alpha^\prime$). 

{\it Borophene bilayer.} The energetically stable borophene bilayer (BBL) structure, predicted by Wu {\it et al.}, is characterized by the AA stacking of BML(${\alpha^\prime}$)s connected by one interlayer B-B bond per periodic surface unit cell, Fig.\,\ref{models0}(a). As discussed in Ref.\,\cite{liuNatMat2021}, the formation of interlayer B--B bonds are mediated by additional electrons resulting from the hollow hexagons on the borophene sheets. In this case,  the densities of interlayer bonds ($u$) and hollow hexagons are the same, $u$\,=\,$\eta$\,=\,1/9, here refereed as \bl. 

\begin{table}
 \centering
  \caption{\label{energy0} Formation energy\,\cite{FormEnergy2} ($E^f$ in eV/B-atom) of borophene bilayer (BBL) and the interlayer vertical B-B bond length (in \AA); $\eta$ and $u$  indicate the densities of hollow hexagons and interlayer B-B bonds. The number of B-B bonds per surface unit cell is indicated in parentheses.}
\begin{ruledtabular}
  \begin{tabular}{lcccc}
 BBL      &   $\eta$    & $u$   &  $E^f$   &  B-B (\AA) \\
\hline
 \bl\,      &   1/9     &  1/9  &   -0.037  &  1.70  \\
 \bla\,     &   1/12    &  1/4  &    -0.149   & 1.69(1)/1.77(2) \\  
  \blb\,     &   1/12    &  1/4  &  -0.153  & 1.72(2)/1.77(1)     \\
%   \blc\,     &   1/12    &  1/4  &  -0.152  & 1.72(2)/1.76(1)    
 \end{tabular}
\end{ruledtabular}
\end{table}

%%%%%%FIG2
  \begin{figure}
    \includegraphics[width=8cm]{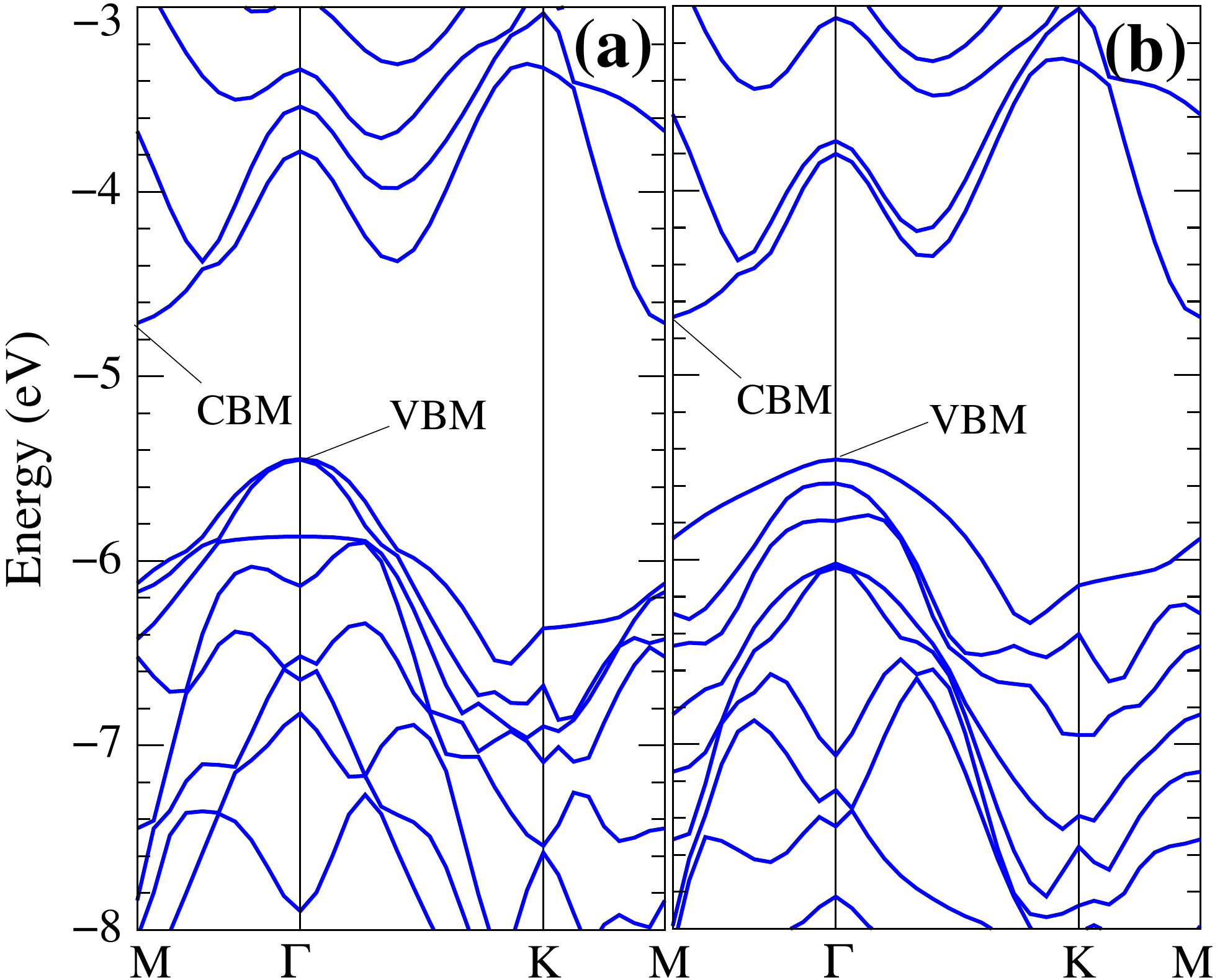}
    \caption{\label{bands0} Electronic band structures of semiconductor borophene bilayer  \bla\,(a) and \blb\,(b); VBM and CBM indicate the valence band maximum and the conduction band minimum. The zero energy was set to the vacuum level.}
  \end{figure}
  
Recent studies, based on first-principles DFT simulations,\cite{maNanoRes2022prediction, xuNanoLett2022quasi} predicted the formation of energetically and structurally stable hexagonal structures with $\eta$\,=\,1/12 and $u$\,=\,1/4, hereafter referred to as \bla\, and \blb,  shown in Figs.\,\ref{models0}(b) and (c), respectively. In \bla\, the BML sheets are eclipsed (A-A stacked), while in \blb\, the borophene layers are staggered (A-B stacking). They found that (i) these two BBL systems present nearly the same total energy, with the latter more stable by 3\,meV/B-atom, and (ii) both are  energetically more stable than \bl\, by about 149\,meV/B-atom. Indeed, as shown in Table~\ref{energy0}, we found  that  the total energy of \blb\, is 4\,meV/B-atom lower in comparison with the  one of  \bla [(i)], and both  are energetically more stable than \bl\, by about 116\,meV/B-atom [(ii)].

%%%%%%FIG3
\begin{figure}
    \centering
    \includegraphics[width=\columnwidth]{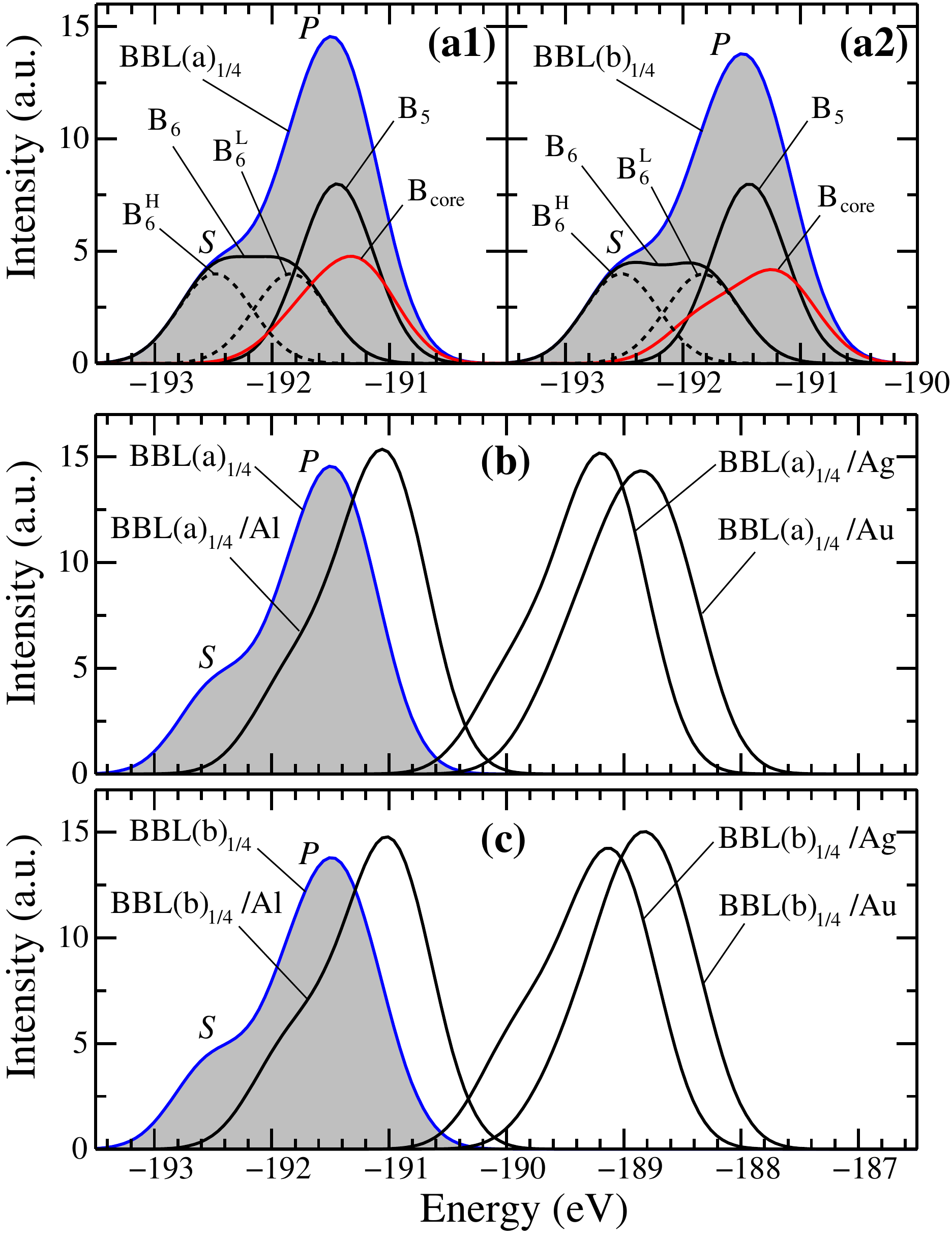}
    \caption{\label{xps} B-1s XPS simulation for the freestanding BBL with (a) {\bla} and (b) {\blb}. The reference level was taken as the experimentally observed for h-BN (190.1\,eV) \cite{PCCPrastogi2019}. We have used a Gaussian width of $0.3$\,eV.}
\end{figure}
 
In Figs.\,\ref{bands0}(a) and (b) we present the electronic band structures of \bla\, and \blb\, systems. These BBL systems are semiconductors, with indirect energy gaps ($E_\text{gap}$) of 0.75 and 0.77\,eV, calculated with the GGA-PBE approach.  We found the valence (conduction) band maximum (minimum) at the $\Gamma$ (M) point, resulting in ionization-potential (IP) and electronic-affinity (EA) of 5.46 and 4.71\,eV [for \bla], and 5.46 and 4.69\,eV [for \blb], which are somewhat comparable with those found in MoS$_2$.\cite{jiang2012electronic, schlaf1999band} 

Further structural characterization was carried out through a set of simulations of the B-$1s$ X-ray photoelectron spectra (XPS). The  similarities in the structural and electronic properties  of  the \bla\, and \blb\, are reflected in their XPS results. Indeed, as shown in Figs.\,\ref{xps}(a1) and (a2) (shaded region), we find that  AA and AB stackings present essentially the same XPS spectrum. However, despite such similarities, it is worth examining the spectral contribution of boron atoms as a function of local geometry and coordination, specifically (i) the spectra B$_5$ of the five-fold coordinated atom [indicated by pentagons in Fig.\,\ref{models0}(b1) and (c1)]  are characterized by a single peak  with a BE of 191.4\,eV, thus,  indicating that these atoms are at the same structural and chemical environments. In contrast, (ii) the B-$1s$ electrons of the six-fold coordinated atoms present higher values of BEs, and their spectra (B$_6$) are composed by a superposition of two sub-peaks [dashed lines in Fig.\,\ref{xps}(a)] with BEs of about 192.5 (B$_6^\text{H}$) and 191.8\,eV (B$_6^\text{L}$), where the one with the higher (lower) BE is characterized by six (four) next-nearest neighbor (NNN) coordination [red (blue) hexagons in Fig.\,\ref{models0}(b1) and (c1)]. (iii) The boron atoms that form the interlayer (vertical) bonds present the lowest values of B-$1s$ BEs, B$_\text{core}$ [red solid line in Fig.\,\ref{xps}(a)].  These core atoms are six- and seven-fold coordinated, as indicated by the blue and red circles in Fig.\,\ref{models0}(b1) and (c1), where it is noticeable a slightly larger BE dispersion of B$_\text{core}$ in \blb\, when compared with that in \bla, Figs.\,\ref{xps}(a2) and (a1), respectively. 

Finally, it is worth noting that BE's peak $P$ at $\sim$191.5\,eV  results from the superposition of B$_5$ (main contribution), B$_\text{core}$ and B$_6^\text{L}$. The five-fold coordinated  atoms, which are aside from the hollow hexagons, make the major contribution to the BE peak $P$, followed by the core atoms that form the interlayer B\,-\,B bonds, and the six-fold coordinated atoms  with four NNN  boron atoms.  The shoulder $S$, meanwhile, is a BE fingerprint of B$_6^\text{H}$ from the six-fold coordinated atoms with six NNN boron atoms.

\subsection{Borophene on the metal (111) surface}

In the sequence,  we have studied the energetic properties, structural characterization, electronic, and electronic-transport properties of the BBL adsorbed on the metal surfaces (BBL/m-surf), namely \bla\, and \blb\, on the Ag(111), Au(111), and Al(111) surfaces, Fig.\,\ref{models0}(d).

\subsubsection{Energetic  and structural properties}

The energetic stability of the BBL/m-surf systems can be determined through the calculation of the binding energy ($E^b$) defined as 
$$ E^b = E[\text{BBL/m-surf}] - E[\text{BBL}] - E[\text{m-surf}], $$ 
where $E[\text{BBL/m-surf}]$ is the total energy of the final system, and $E[\text{BBL}]$ and $E[\text{m-surf}]$ are the total energies of the isolated components, borophene bilayer, and metal surface. We have examined  three interface configurations for \bla/Ag(111), namely, with the \bla\, hexagonal hollow site aligned with (i) bridge, (ii) hollow, and (iii) on-top sites of the Ag(111) surface. We found that (i) is energetically more favorable than (ii) and (iii)  by  1.6 and  4.0\,meV/\AA$^2$, respectively. We have considered the same interface geometry [(i)] for the other BBL/m-surf systems.

\begin{table}
 \centering
  \caption{\label{energy1} Binding energy ($E^b$ in meV/\AA$^2$) of the BBL/metal-(111) interface, the equilibrium vertical distance between the BBL and the metal(111) surface ($h$), and the interlayer B-B vertical bond length.  Within the parenthesis, it is stated how many vertical B-B bonds are present in each surface unit cell.}
\begin{ruledtabular}
  \begin{tabular}{lccc}
  BBL/metal(111)      &  $E^b$ &   $h$ (\AA)   &  B-B (\AA) \\ 
\hline
 \underline{BBL/Ag(111):}   &         &           &                        
    \\
\bla           & $-28$      & 2.508  &  1.70(1)/1.79(2)   \\
\blb           & $-31$      &   2.427    &   1.74(2)/1.79(1)   \\
    
\hline
\underline{BBL/Au(111):}       &         &           &                     
  \\
\bla           &  $-28$  &      2.272           &  1.71(1)/1.79(2)   \\
\blb           &  $-39$  &      2.267           &  1.74(2)/1.78(1)   \\
\blb (new2)    &  $-33$  &      2.250           &  1.74(2)/1.78(1)   \\
\hline
\underline{BBL/Al(111):}   &            &           &          \\
\bla             & $-33$   &      2.428         & 1.71(1)/1.79(2)      \\
\blb             & $-34$   &       2.441        & 1.74(2)/1.79(1) 
 \end{tabular}
\end{ruledtabular}
\end{table}

\begin{table}
\centering
  \caption{\label{Drho} Work function ($\Phi$) and the  charge transfer, $\Delta\rho$,  at the BBL/me-surf interface. Positive values of $\Delta\rho$ indicate a net electron transfer from the metal surface to the  borophene bilayer. }
\begin{ruledtabular}
  \begin{tabular}{lll}
  BL/metal(111)           &  $\Phi$\,(eV)   & $\Delta\rho$\,($e/\text{cm}^2$)\\ 
\hline
\underline{BL/Ag(111):}  &           &         \\             
\bla                      &  4.777    & $+5.1\times 10^{13}$\\
\blb                      &  4.680    & $+5.0\times 10^{13}$\\   
\hline
\underline{BL/Au(111):}   &           &      \\           
\bla                      &   4.632   & $-5.3\times 10^{13}$ \\
\blb                      &   4.680    & $-5.0\times 10^{13}$    \\
\blb\, (new2)             &   4.658    & $-5.7\times 10^{13}$  \\
\hline
\underline{BL/Al(111):}   &           &            \\
\bla                      & 4.760     & $+4.5\times 10^{14}$ \\
\blb                      & 4.721     & $+4.6\times 10^{14}$ 
\end{tabular}
\end{ruledtabular}
\end{table}

%%%%%%FIG4
  \begin{figure}
    \includegraphics[width=8cm]{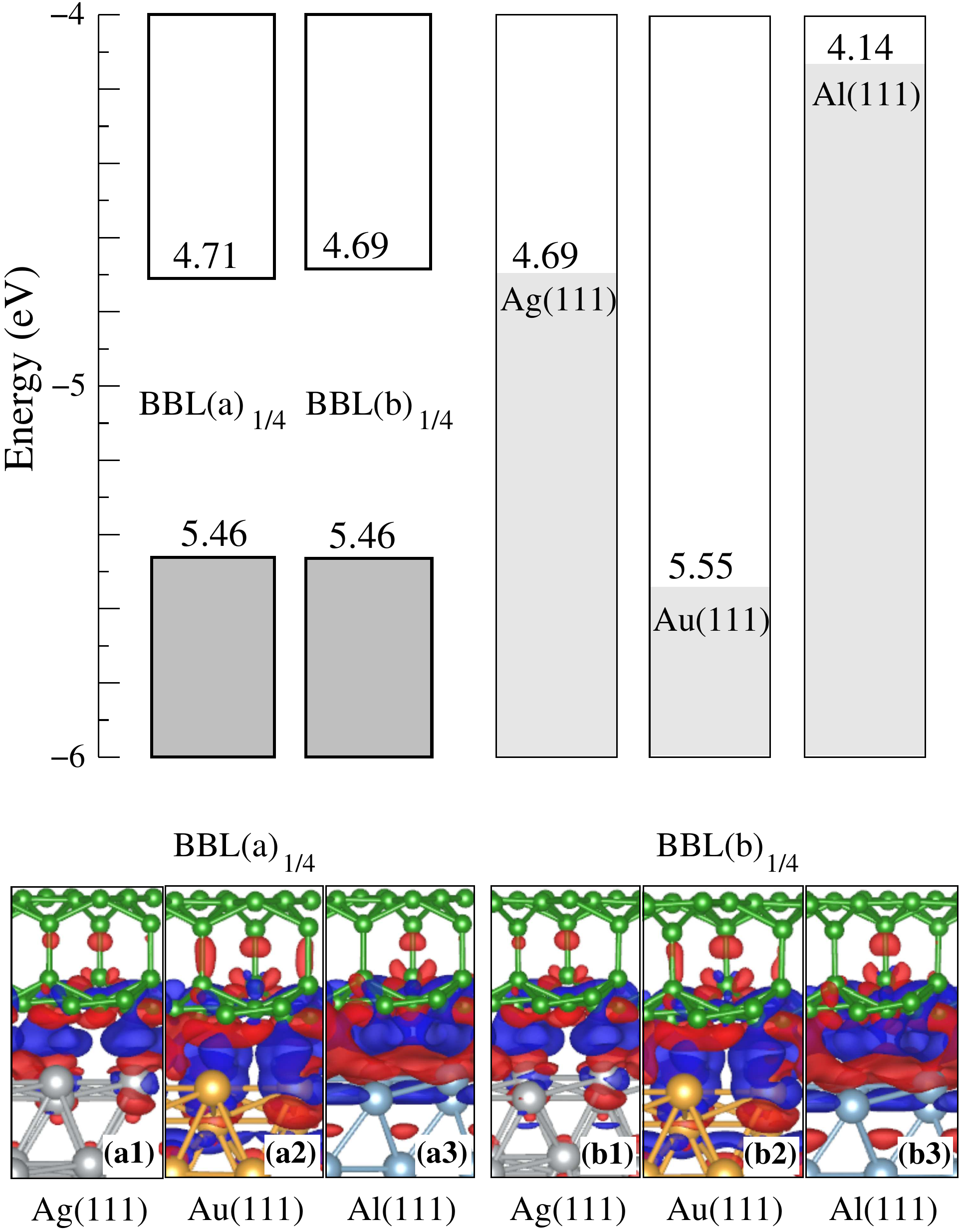}
    \caption{\label{levels}\,(top) Ionization potential and electronic affinity of \bla, and \blb, and the work-function of pristine Ag(111), Au(111), and Al(111) surfaces in eV. Net electron transfers, $\Delta\rho$, at the \bla/m-surf (a1)-(a3), and \blb/m-surf (b1)-(b3), with m-surf\,=\,Ag, Au and Al (111) surfaces. Blue (Red) isosurfaces indicate $\Delta\rho>0$ ($\Delta\rho<0)$, with isosurface of 0.001\,$e$/\AA$^3$.}
  \end{figure}
  
Our binding energy results (Table~\ref{energy1}) are comparable with those predicted for other 2D systems, for instance MoS$_2$/Ag(111),\cite{gong2014unusual,zhong2015interfacial}  and  can be classified as “potentially exfoliable”.\cite{mounetNatNanotech2018} Indeed, such an exfoliable nature of \blb\, adsorbed on Ag(111) was pointed out (very recently) by Xu {\it et al.}\cite{xuNanoLett2022quasi}, where they found $E^b$ of $-107$\,meV/\AA$^2$ using the Grimme formula (vdW-D2), and Li {\it et al.}\cite{li2022opposite} where they obtained  $E^b$ of 100\,meV/\AA$^2$ and $h$=2.26\,\AA\, using vdW-D3.\cite{ehrlich2011system} Here,  using the same vdW-D2 approach we found a binding energy of $-111$\,meV/\AA$^2$. In Table\,\ref{energy-ap} (Appendix) we present our results of binding energy and equilibrium geometry of \blb/Ag(111) obtained using other vdW approaches.

%%%%%%FIG5  
  \begin{figure*}
    \includegraphics[width=15cm]{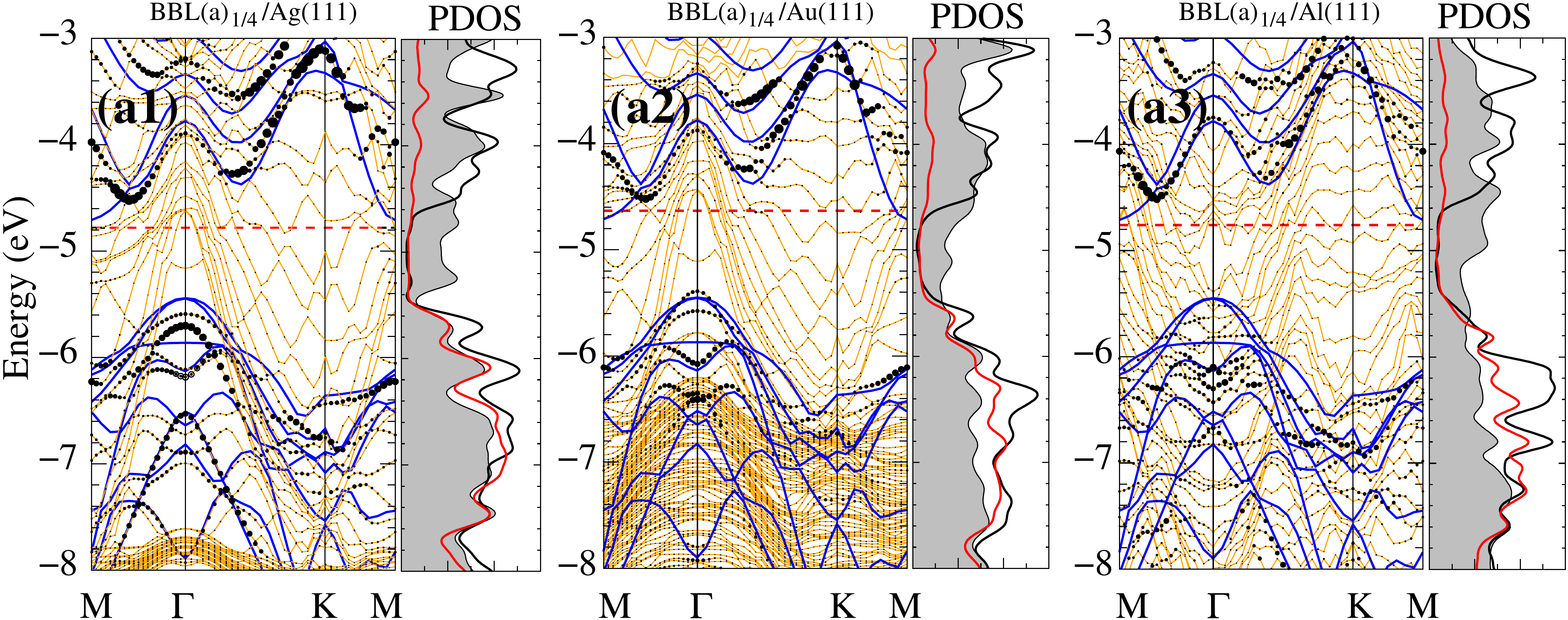}
    \includegraphics[width=15cm]{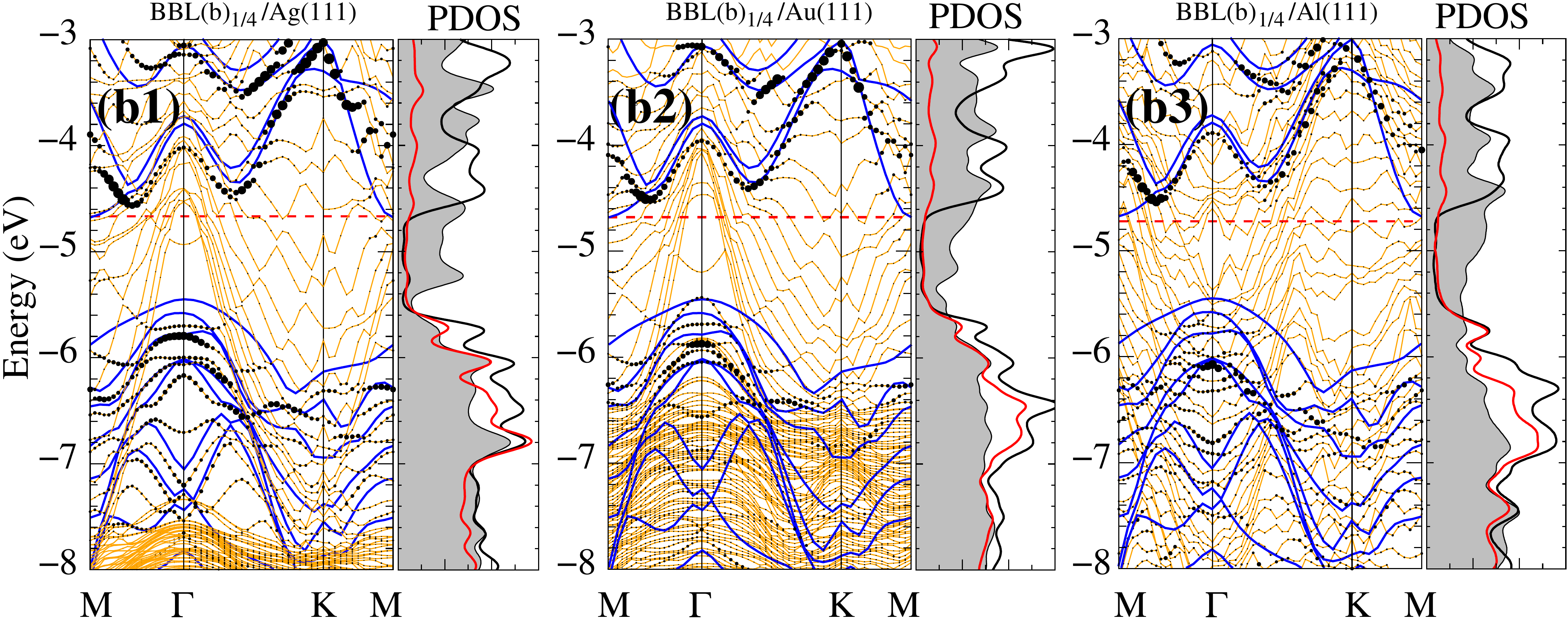}
    \caption{\label{fatbands} Electronic band structures and the B-2$p$ projected density of states (PDOS) of \bla\,(a1)-(a3), and \blb\,(b1)-(b3) on the Ag (a1)/(b1), Au (a2)/(b2), and Al (a3)/(b3) (111) surface. Filled black circles (fat bands) are proportional to the PDOS of the $2p_z$ orbitals of the topmost boron atoms. The electronic band structures of pristine BL are shown overlayed in solid blue lines, with orange lines the full system bands [BBL/m-surf], and the red-dashed lines indicate the Fermi level. PDOS of the bottom-most (shaded region), central (red solid-lines), and topmost (black solid-lines) B-2$p$ orbitals. The zero energy was set to the vacuum level.}
  \end{figure*}

\subsubsection{Electron/Hole doping}
  
%%%%%%FIG6
  \begin{figure*}
    \vspace{5mm}
    \includegraphics[width=14cm]{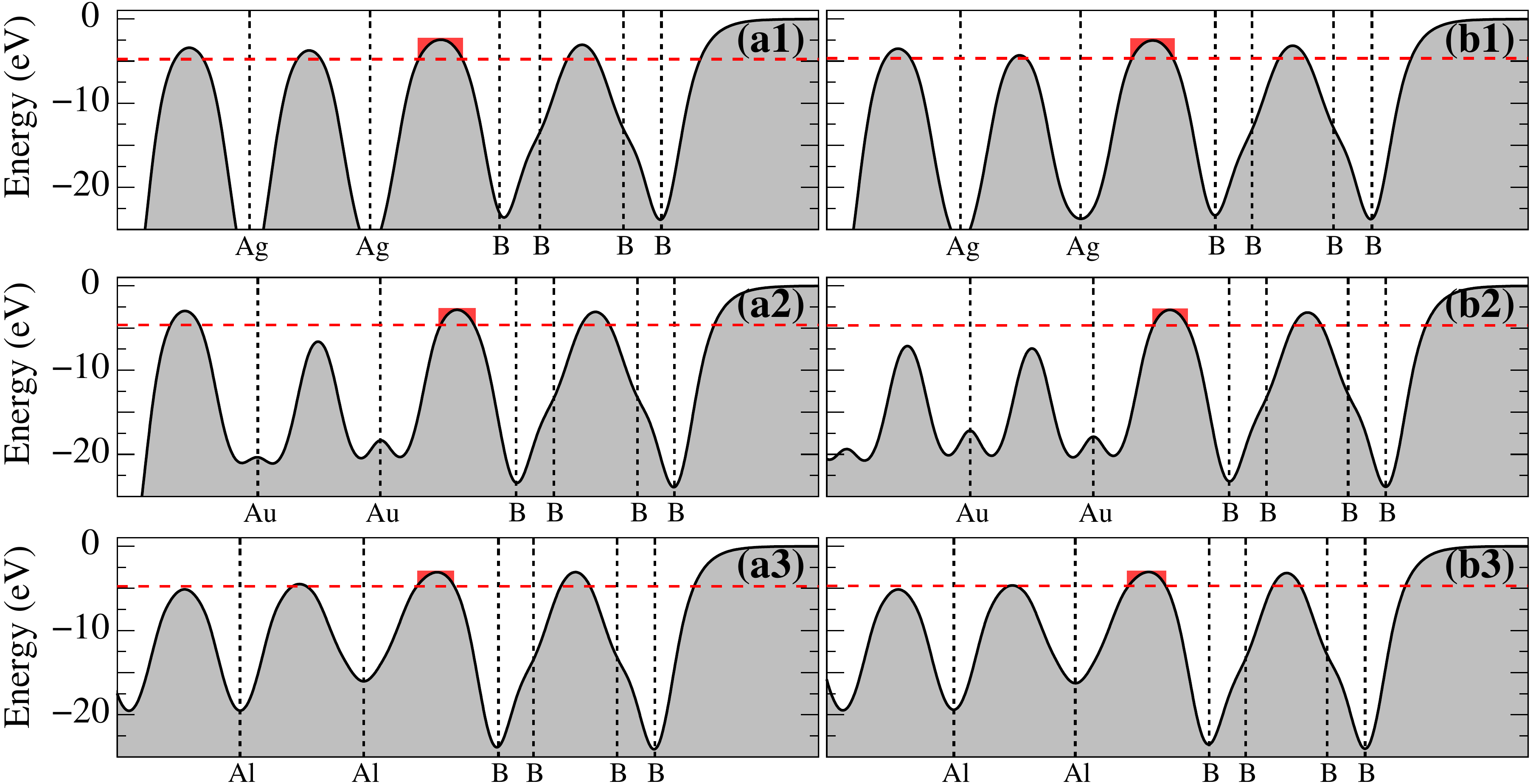}
    \caption{\label{avg} Planar averaged electrostatic potential along the \bla/m-surf  (a1)-(a3), and \blb/m-surf (b1)-(b3), with m-surf\,=\,Ag(111), Au(111), Al(111) surfaces. Red dashed lines indicate the energy position of the Fermi level, and the rectangular shaded regions indicate the energy barrier used for the calculation of $T_\text{B}$.}
  \end{figure*}

The energy positions of the ionization potentials (IP) and electronic affinity (EA) of free-standing borophene bilayers, as well as the work function of the pristine (111) metal surfaces ($\Phi_\text{m-surf}$), are shown in Fig.\,\ref{levels} (top-diagram). Our findings of $\Phi_\text{m-surf}$ are in good agreement with the experimental results, \cite{derry2015recommended}  viz.: $\Phi_\text{Ag}$\,=\,4.53;  $\Phi_\text{Au}$\,=\,5.33; $\Phi_\text{Al}$\,=\,4.32\,eV [Table~\ref{Drho}]. 

When the BBL/m-surf interface forms, there is a net electron transfer from the Ag(111) and Al(111) surfaces to the BBLs, $n$-type doping, whereas in BBL/Au(111) the electron transfer occurs in the opposite direction, and the BBLs become $p$-type doped. The $n$- and $p$-type doping level of the BBLs were examined  using the Bader charge density analysis,\cite{bader}  see Table\,\ref{Drho}. Meanwhile, the net charge transfer profile ($\Delta\rho=\rho_\text{BBL/m-surf}-\rho_\text{BBL}-\rho_\text{m-surf}$) along the BBL/m-surf systems, shown in Figs.\,\ref{levels}(a) and (b), reveal that the charge transfers [$\Delta\rho>0$ (blue), and $\Delta\rho<0$ (red)] are localized in the BBL/m-surf interface region.

%%%%%%FIG7
  \begin{figure}
    \vspace{5mm}
    \includegraphics[width=7.5cm]{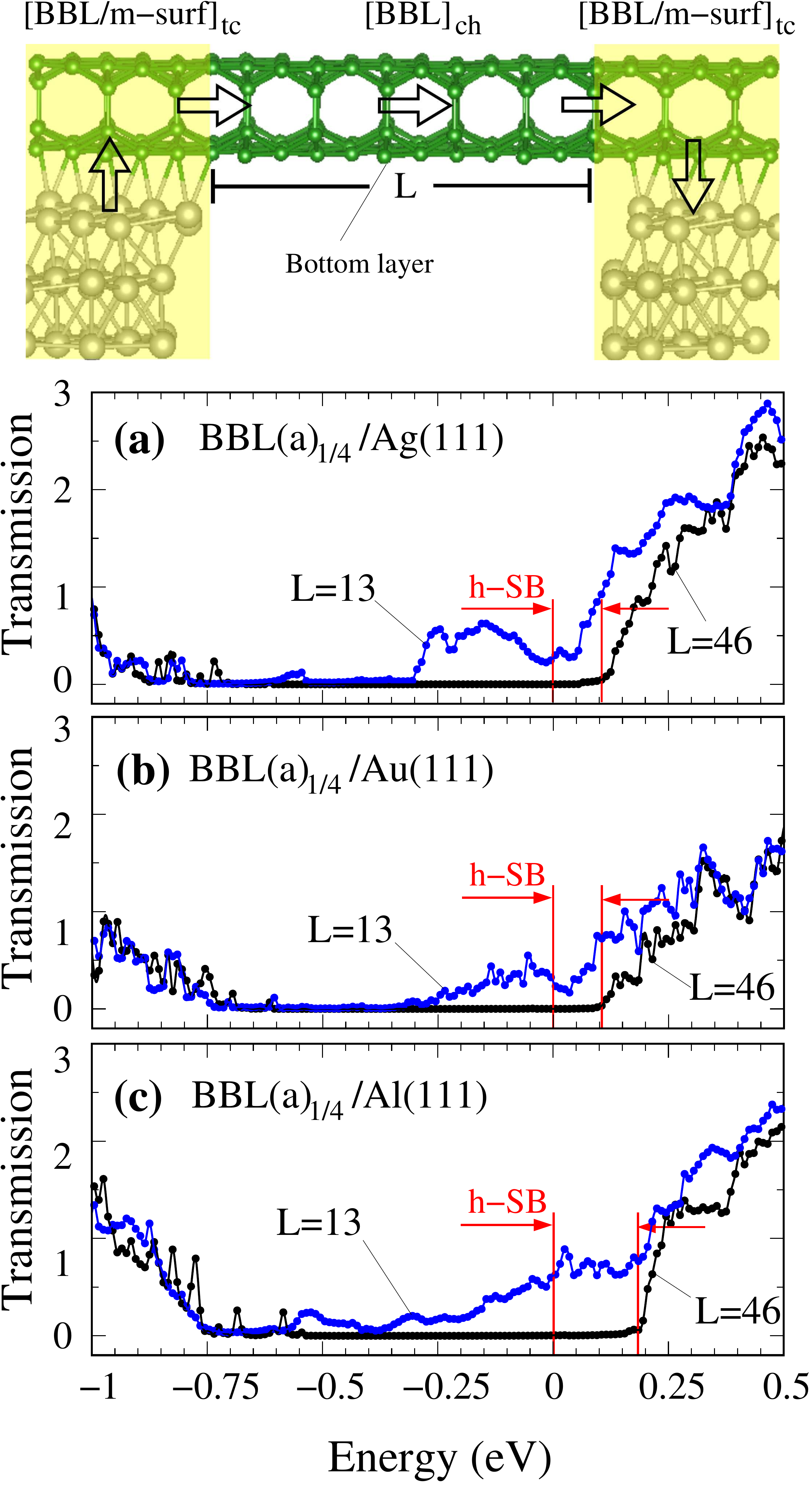}
    \caption{\label{trans} (top) Schematic model of top-contact, [BBL/m-surf]$_\text{tc}$, and the BBL channel, [BBL]$_\text{ch}$,  [BBL/m-surf]$_\text{tc}$\,-\,[BBL]$_\text{ch}$\,-\,[BBL/m-surf]$_\text{tc}$. Transmission probabillity through [\bla]$_\text{ch}$ with  m-surf top-contact of (a) Ag(111), (b) Au(111), and (c) Al(111) for channel lengts of L\,=\, 13 and 46\,\AA.  The zero energy was set to the Fermi level.}
  \end{figure}

\subsubsection{Structural characterization: XPS}

We found that the atomic lattice structures of the BBLs have been retained even though chemical bonds have formed at the BBL/m-surf interface. Indeed, our simulated XPS spectra of the BBL/m-surf systems [Figs.\,\ref{xps}(b) and (c)] reveal nearly the same XPS features as those obtained in the free-standing BBLs [Fig.\,\ref{xps}(a)]. However, characterized by a reduction of the B-$1s$ BEs, giving rise to negative values of (B-$1s$) core level shifts (CLSs), $\Delta\text{BE}$. For instance, in \bla/Ag(111) [Fig.\,\ref{xps}(b)], it is noticeable  the reduction of the BE associated with the more intense spectra, BE\,=\,189.2 eV, resulting in a  CLS of $\Delta\text{BE}=-2.3$\,eV when compared with the energy position of its counterpart (peak $P$) of pristine \bla\, [Fig.\,\ref{xps}(a1)], and  the almost  suppression of the shoulder $S$ attributed to the BE spectra  B$_6^\text{H}$. These features  were also observed in the other BBL/m-surf systems, with $\Delta\text{BE}$ of  about $-2.7$ and $-0.5$ for BBL/Au(111) and /Al(111), Fig.\,\ref{xps}(b) and (c), independent of the BBL geometry. 

The lowering of the B-$1s$ core-level BE, $\Delta\text{BE}<0$, can be qualitatively understood in light of initial state mechanisms, namely (i) net charge transfers, (ii)  orbital hybridizations, and (iii) emergence of the external electric fields (EEFs) at the BBL/metal interface.\cite{JESRPbagus1999,pueyo2015prediction} Our results indicate that the former two mechanisms play a minor role. For example, the net electron transfers, based on the Bader analysis\cite{bader} (Table\,\ref{Drho}) we find that (on average) each boron atom of \bla/m-surf is electron-doped by  about 0.007 and 0.014\,$e$ for m-surf\,=\,Ag(111) and Al(111), respectively, whereas in \bla/Au(111) the boron atoms are hole-doped by  $\sim$\,0.007\,$e$/B-atom. We estimate the B-$1s$ CLSs  induced by such amount of charge transfers (Appendix, see Table\,\ref{cls-free}). The BE of the B$_5$ spectrum is found to be reduced by $-0.14$ and $-0.28$\,eV in the $n$-type doped \bla/Ag(111), and /Al(111), whereas in the $p$-type doped \bla/Au(111), the B$-1s$ BE increases by $\Delta\text{BE}$\,=\,$+0.15$\,eV. In (ii), the L\"{o}wdin charge density analysis\cite{lowdinJChemPhys1953} reveals that the formation of the BBL/metal interfaces is characterized by a slight increase in the electron occupation of BBL/metal interface B-$2p$ orbitals [$\Delta\rho(\text{B}\text{-}2p)$] in comparison with those of free-standing BBLs. For example,  we found $\Delta\rho(\text{B}\text{-}2p)$ of $\sim$0.12\,$e$/B-atom for the five-fold boron atoms at the \bla/metal interface. These findings suggest that the screening effects induced by the EEFs [(iii)] make the main contribution to the B-$1s$ CLSs. The emergence of such EEFs is due to the formation of electron- and hole-rich centers localized at the BBL/metal interface, blue and red isosurfaces in Figs.\,\ref{levels}(a) and (b), where the negative values of CLSs, $\Delta\text{BE}<0$, indicate the predominance of electron-repulsive potentials ($\Delta\rho>0$) on the B-$1s$ core levels. 

%%%%%%FIG8
  \begin{figure}
    \vspace{5mm}
    \includegraphics[width=7.5cm]{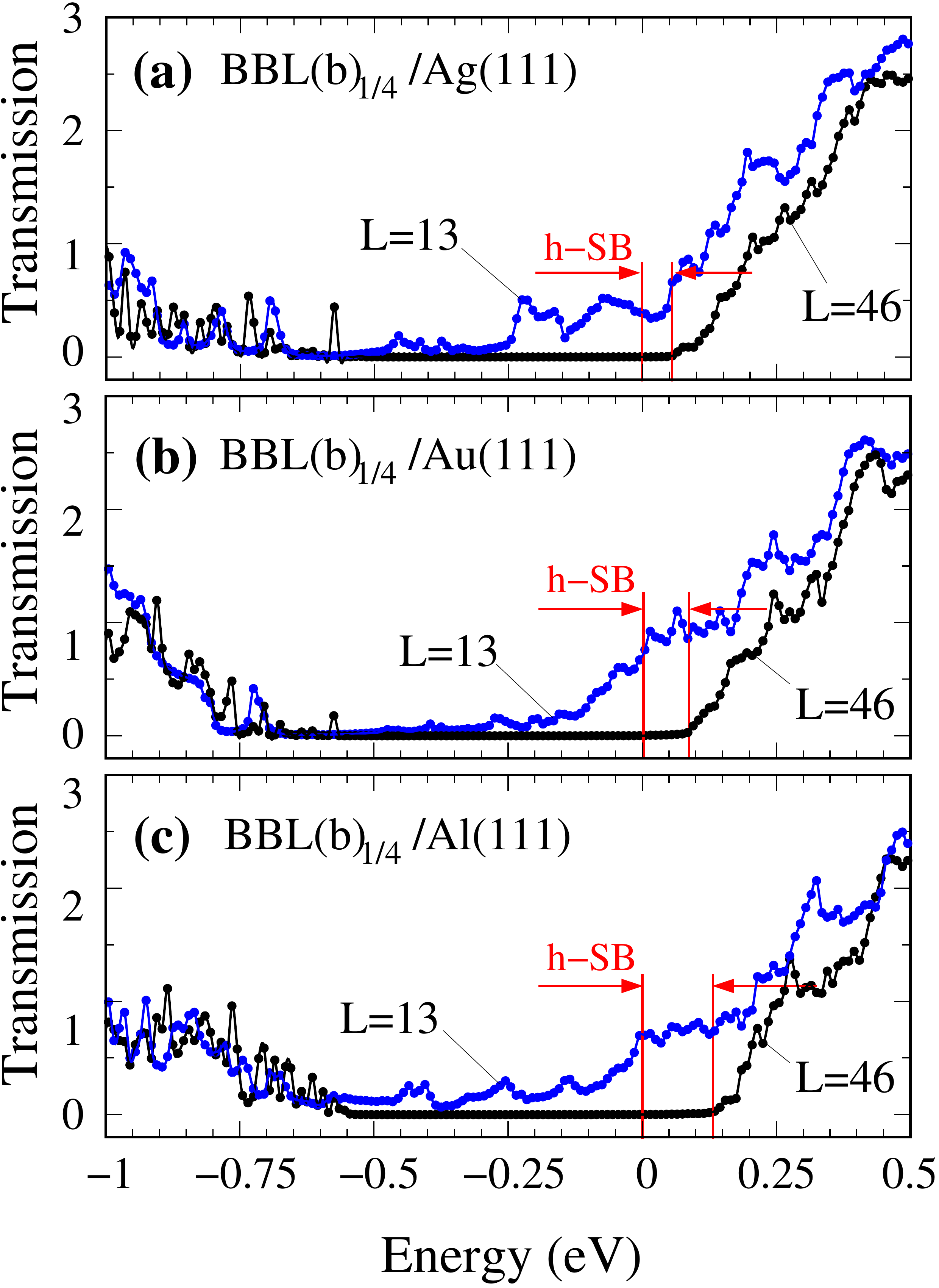}
    \caption{\label{transb}  Transmission probabillity through \blb\, channel with  metal top-contact of (a) Ag(111), (b) Au(111), and (c) Al(111) for channel lengts of L\,=\, 13 and 46\,\AA. The zero energy was set to the Fermi level.}
  \end{figure}

\subsubsection{Electronic band structure}

In Figs.\,\ref{fatbands}(a) and (b) we show the electronic band structure  of \bla/ and \blb/m-surf, where the electronic band structures of pristine \bla\, and \blb\, are indicated in blue solid lines. It is noticeable (i) the formation of metallic bands within the energy gap of the pristine BBLs, and (ii) the absence of localized interface states. The projected density of states (PDOS) on the bottom-most boron atoms (shaded region in Fig.\,\ref{fatbands}) reveal (iii) the emergence of orbital overlaps at the  BBL/metal interface, which, in combination with (i) and (ii),  may lead to a lowering of the BBL/metal contact resistance. As a result of such orbital overlap  and the BBL\,$\leftrightarrow$\,m-surf net charge transfer, the Fermi level (red dashed lines) is pinned near the CBM of pristine BBLs (blue solid lines);  resulting in work functions ($\Phi$) between 4.6 and 4.8\,eV, almost independent of the metal surface [Table\,\ref{Drho}].  In contrast, the PDOS of the inner and topmost layers of  \bla\, and \blb, red and black solid lines in Fig.\,\ref{fatbands}, reveal that the semiconducting character is preserved in the BBL's inner and top layers. The orbital projected band structure on the topmost boron layer (filled circles in Fig.\,\ref{fatbands}) indicates that the CBM lies along the M-$\Gamma$ direction, which results in a  $n$-type Schottky barrier between 0.1 and 0.3 eV.  It is important to note that such Schottky barriers are with respect to the topmost and inner layers of BBL; when we consider the bottom-most layer, there is no Schottky barrier; instead, we have a BBL/m-surf ohmic contact due to the metalization of the interface states.

\begin{table}
 \centering
  \caption{\label{tunnel} Tunnelling barrier height ($\Delta V$ in eV), width ($w_B$ in \AA) and tunnelling probability ($T_B$) for the BBL/m-surf systems.}
\begin{ruledtabular}
  \begin{tabular}{lccc}
  BL/metal(111)      &  $E_0$ & $w_\text{B}$ &  $T_\text{B}$  \\ 
\hline
 \underline{BL/Ag(111):}   &       &              & \\             
\bla                 &  2.31       & 0.90         & 0.25 \\
\blb                 &  2.12       & 0.87         & 0.27 \\
\hline
\underline{BL/Au(111):}  &         &         & \\  
\bla                 &  1.79       & 0.73    & 0.37 \\
\blb                 &  1.83       & 0.70    & 0.38\\
\hline
\underline{BL/Al(111):}    &      &         & \\
\bla                &   1.67      &   0.80  & 0.35 \\
\blb                & 1.66        &   0.80  & 0.35\\
 \end{tabular}
\end{ruledtabular}
\end{table}

\subsubsection{Metal\,$\rightarrow$\,BBL electron tunneling barrier}

The maintenance of the semiconducting properties in the boron topmost and core layers combined with the formation of metallic channels at the BBL/metal contact region, as reported in (i)-(iii) above,
% is somewhat similar to the partial Fermi level pinning predicted in MoS$_2$/metal interfaces,\cite{gong2014unusual}
suggest that the BBL/m-surf is a promising platform to be used in the design of electronic devices based on 2D materials. In that sense, the tunneling barrier at the BBL/m-surf interface is an important feature in order to evaluate the electron injection efficiency at the contact region. The averaged electrostatic  potential  along the BBL/m-surf system (Fig.\,\ref{avg}) is characterized by the rise of a tunneling barrier at the interface region.  The tunneling probability ($T_{\text{B}}$) is ruled by the height ($E$) and the width ($w_\text{B}$) of the tunneling barrier.   Assuming that the tunneling barrier height is constant ($E_0$) with a rectangular shape and width equal $w$, as indicated by the red shaded  regions in Fig.\,\ref{avg}, we can infer  $T_\text{B}$ as,
$$
T_\text{B}=\text{exp}\left(-2\frac{\sqrt{2mE_0}}{\hbar}\times 
w_\text{B}\right),
$$
within the WKB approximation.\cite{pan2016monolayer, liu2018tis, liu2021tuning} Our results of $T_\text{B}$, presented in Table\,\ref{tunnel},  are comparatively larger than those obtained in septuple-layered MoSi$_2$N$_4$ and WSi$_2$N$_4$ MLs,\cite{wang2021efficient}  and MoS$_2$ ML and BL\cite{zhong2015interfacial} on Au and Ag surfaces, however smaller than those predicted for phosphorene ML on the same metal surfaces.\cite{pan2016monolayer}

%%%%%%FIG9
  \begin{figure}
    \vspace{5mm}
    \includegraphics[width=8.5cm]{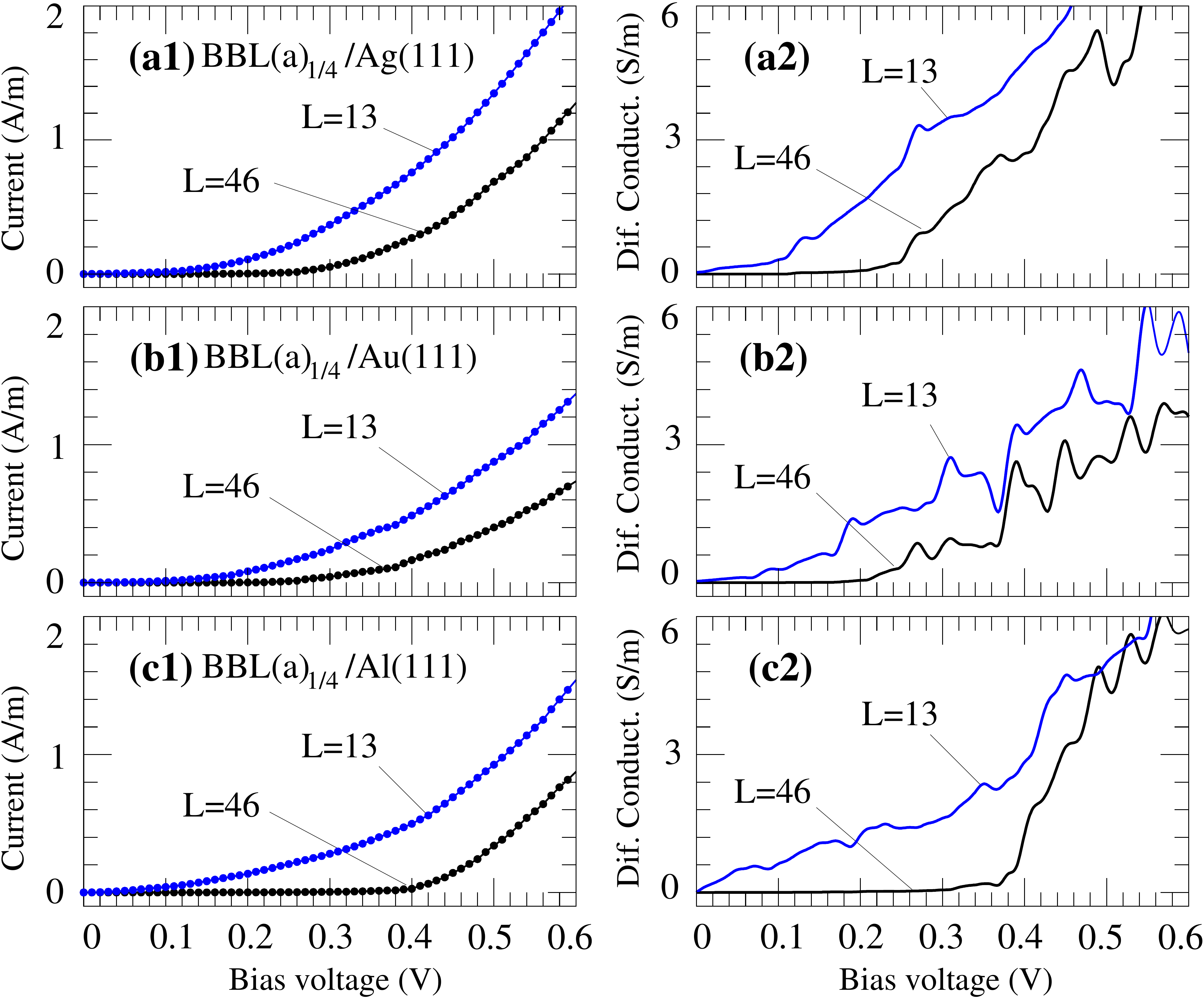}
    \caption{\label{current} Electronic current ($I$) (a1)-(c1) and differential-coductance ($dI/dV$) (a2)-(c2) of [\bla/m-surf]$_\text{tc}$\,-\,[\bla]$_\text{ch}$\,-\,[\bla/m-surf]$_\text{tc}$ with m-surf\,=\,Ag(111), Au(111) and Al(111). L indicates the \bla\, channel length in \AA.}
  \end{figure}
\subsubsection{Electronic transport along the BBL channels}

In order to provide a complete picture of BBL as a viable and promising material for ``two-dimensional electronics'', we calculate the electronic transmission function, $Tr(E)$ [eq.\,(1)], taking into account the BBL/m-surf top contact (tc) region connected with the semiconductor BBL channel (ch), [BBL/m-surf]$_\text{tc}$\,-\,[BBL]$_\text{ch}$\,-\,[BBL/m-surf]$_\text{tc}$ schematically shown in the top diagram of Fig.\,\ref{trans}. Our results of $Tr(E)$ are depicted in Figs.\,\ref{trans}(a)-(c) and \ref{transb} for \bla/ and \blb/m-surf contacts, respectively, where we have considered BBL channel lengths (L) of 13 and 46\,\AA. 

For L\,=\,13\,\AA, we find transmission features with $Tr(E)>0$ around the Fermi level, which can be attributed to orbital overlaps  (predominantly) along the bottom boron layer of  [BBL]$_\text{ch}$ connecting the left and right contacts, [BBL/m-surf]$_\text{tc}$. By increasing the length of the BBL channel to L\,=\,46\,\AA\, such an orbital overlap is suppressed, giving rise to transmission gaps, $Tr(E)$\,=\,0, around the Fermi level. For instance, in \bla/Ag(111) [Fig.\,\ref{trans}(a)] we find $Tr(E)$\,=\,0 for $|E-E_F|\,\lesssim\,0.1$\,eV. Meanwhile, increasing the energy window towards positive values,  $E-E_F\,\gtrsim\,0.1$\,eV, we will find the emergence of  electronic transport ($Tr(E)>0$) mediated by the conduction band of [\bla]$_\text{ch}$. As a result, we find a $n$-type horizontal Schottky barrier (h-SB) of about 0.1\,eV between  [\bla/Ag(111)]$_\text{tc}$ contact and [\bla]$_\text{ch}$ channel. 

It is worth noting that the transmission gap extends to an energy interval of $\sim$\,0.7\,eV below the Fermi level, i.e.  $Tr(E)$\,=\,0 for $-0.7\,\text{eV}\lesssim\,E-E_F\lesssim\,0.1\,\text{eV}$, resulting in a total transmission gap of about 0.8\,eV which is very close to the band-gap of the free-standing \bla. That is, even though orbital hybridizations (metallizations) take place at the \bla/Ag(111) contact interface, the bandgap of \bla\, (channel) is recovered for L$\gtrsim$\,5\,nm.

The electronic transport results in the  other systems, Figs.\,\ref{trans}(b)-(c) and \ref{transb}(a)-(c), are somewhat comparable with those obtained for [\bla/Ag(111)]$_\text{tc}$\,-\,[\bla]$_\text{ch}$\,-\,[\bla/Ag(111)]$_\text{tc}$ [Fig.\,\ref{trans}(a)], characterized by the formation of an $n$-type h-SB barriers between 0.1 and 0.2\,eV almost independent of the BBL geometry. Indeed, this is in consonance with the fact that BBL/m-surf systems present practically the same work function $\Phi$ (between 4.6  and 4.8\,eV) and \bla\, and \blb\, have nearly identical ionization potentials and electron affinity, Fig.\,\ref{levels}. 

By using the Landauer-B\"uttiker formula [eq.\,(2)], we examined the electronic current ($I$) and the respective differential-conductance ($dI/dV$) through the \bla\, channels with L\,=\,13 and 46\,\AA.  We have considered the limit of low bias voltage, which allows us to estimate the electronic current using the transmission probability at zero-bias. Our results, presented in Fig.\,\ref{current}, show that at the short-channel limit, L\,=\,13\,\AA, the absence of horizontal Schottky barrier is characterized by the non-zero values of differential conductances, $dI/dV$\,$>$\,0. In contrast, for a channel length of L\,=\,46\,\AA\, the h-SBs are identified by the voltage intervals ($\Delta V$) with $dI/dV$\,$\approx$\,0. For example, in the \bla/Ag(111) contact, Fig.\,\ref{current}(a2), we find $dI/dV$\,$\approx$\,0 for $V$ up to 0.2\,eV, which corresponds to $\Delta V\pm 0.1$\,eV with respect to the Fermi level, resulting in an h-SB of  0.1\,eV between the \bla/Ag(111) contact and the \bla\, channel. A similar result was obtained in the \bla/Au(111) contact [Fig.\,\ref{current}(b2)], whereas the h-SB increases to about 0.2\,eV when the aluminum contact, \bla/Al(111), is considered [Fig.\,\ref{current}(c2)]. Finally, it is worth mentioning  that,  when the bias voltage exceeds  the h-SB  ($|\Delta V|>\text{h-SB}$), the aluminum and silver contacts will promote higher values of differential conductivity, $dI/dV$, in comparison with that of the gold contact.

\section{Summary and Conclusions}

By means of  first-principles DFT calculations, we performed a thorough investigation of the electronic and structural properties of borophene bilayer (BBL) in contact with metal surfaces, BBL/m-surf. The formation of BBL/m-surf [with m-surf\,=\, Ag, Au, and Al (111) surfaces] is exothermic with binding energies between 30 and 40\,meV/\AA$^2$, which are comparable with other 2D systems adsorbed on metal surfaces, like MoS$_2$/Ag(111). At the equilibrium geometry, the orbital overlap at the BBL/m-surf interface promote the formation of ohmic BBL/metal contact, and the BBL\,$\leftrightarrow$\,m-surf net charge transfers result in $n$-type doped BBLs for m-surf\,=\,Ag and Al (111) surfaces,  and $p$-type doped  for BBL/Au(111). Further structural and electronic characterization of the BBL/m-surf interfaces  was performed by a set of simulations of the B-$1s$ X-ray photoelectron spectra (XPS), with the goal of providing support and atomic scale understanding for  (future) experimental XPS investigations. Here, we found  that the formation of BBL/m-surf is characterized by a  red shift of the B-$1s$ core level spectra  when compared with the free-standing BBL.  The electronic transport through BBL channels connected to the BBL/m-surf contacts reveals the emergence of a horizontal $n$-type Schottky barrier between 0.1 and 0.2\,eV for channels lengths larger than 5\,nm.  These findings show that borophene bilayers can be an important player in the current research in two-dimensional electronics.

\begin{acknowledgments}

The authors acknowledge financial support from the Brazilian agencies  CNPq, INCT-Nanocarbono, INCT-Materials Informatics, FAPEMIG, FAPES(TO-1043/2022), and the LNCC (SCAFMat2), CENAPAD-SP for computer time.

\end{acknowledgments}

\begin{table}
 \centering
  \caption{\label{energy-ap} Binding energy ($E^b$ in meV/\AA$^2$) of \blb\, on the Ag(111) surface, equilibrium vertical distance ($h$) and the interlayer B-B vertical bond lengths using four different vdW approaches. Within the parenthesis, it is stated how many vertical B-B bonds are present in each unit cell.}
\begin{ruledtabular}
  \begin{tabular}{lccc}
  \blb/Ag(111)    & $E^b$ & $h$ (\AA) & B-B (\AA) \\
\hline  
vdW-DF      &  $-31$     &   2.427    &   1.74(2)/1.79(1)   \\
\hline 
vdW-Opt88   &   $-42$     &   2.359    &    1.74(2)/1.79(1)  \\
vdW-DF-cx   & $ -85$     &   2.259    &    1.73(2)/1.78(1) \\  
vdW-D2      &  $-111 $    &   2.223    &    1.73(2)/1.76(1)
 \end{tabular}
\end{ruledtabular}
\end{table}
\section{Appendix}

\subsection{Binding energies using other vdW approaches}

In Table\,\ref{energy-ap} we present the binding energies of \blb\, adsorbed on Ag(111), and \blb-Ag(111) equilibrium vertical distance ($h$) by using other vdW approaches, namely vdW-Opt88,\cite{klimesJPhysC2010} -cx,\cite{berland2014exchange, berland2014exchange} and D2.\cite{grimmeJCompChem2006}

\subsection{B-$1s$ CLSs of electron/hole doped \bla}

In Table\,\ref{cls-free} we present the calculated B-$1s$ CLSs of charged ($q$) free-standing \bla\, with respect to neutral system, $\Delta\text{BE}=\text{BE}[q] - \text{BE}[0]$, where we found that the B-$1s$ core level BE reduces (increases) upon electron (hole) doping.

\begin{table}
 \centering
  \caption{\label{cls-free} Core level shifts of the $n$- and $p$-type doped (by $q$) free-standing \bla.}
\begin{ruledtabular}
  \begin{tabular}{ccc}
\multicolumn{1}{c}{${q}$} & \multicolumn{2}{c}{$\Delta\text{BE}$ (eV)} \\
 $e$/B-atom ($\times 10^{13}\,e/\text{cm}^2)$       & $n$-type & $p$-type \\
 \cline{1-1}
\cline{2-3}
 0.007 ~ (5.1 )   &  $-0.14$ & $+0.15$  \\
 0.014 ~ (10.2)   &  $-0.28$ & $+0.30$
\end{tabular}
\end{ruledtabular}
\end{table}

\bibliography{bib}%

\begin{thebibliography}{68}
\bibitem [{For()}]{FormEnergy2}%
  \BibitemOpen
  \href@noop {} {}\bibinfo {note} {$E^f$ is defined as the the total energy
  difference between the BBL (BBL$\text{(a)}_{1/4}$ or BBL$\text{(b)}_{1/4}$)
  and the BML($\alpha^\prime$), $E^f=(E[\text{BBL}]-2\times
  E[\text{BML}(\alpha^\prime)])/N_\text{B}$, where $N_\text{B}$ is the number
  of boron atoms per unit cell.}\BibitemShut {Stop}%
\end{thebibliography}

\end{document}